\documentclass[pdflatex,sn-mathphys-num]{sn-jnl}% Math and Physical Sciences Numbered Reference Style
\usepackage{graphicx}%
\usepackage{multirow}%
\usepackage{amsmath,amssymb,amsfonts}%
\usepackage{amsthm}%
\usepackage{mathrsfs}%
\usepackage[title]{appendix}%
\usepackage{xcolor}%
\usepackage{textcomp}%
\usepackage{manyfoot}%
\usepackage{booktabs}%
\usepackage{algorithm}%
\usepackage{algorithmicx}%
\usepackage{algpseudocode}%
\usepackage{listings}%
\theoremstyle{thmstyleone}%
\theoremstyle{thmstyletwo}%

\theoremstyle{thmstylethree}%

\begin{document}

\title[A Theory of Quantum Jumps and its Applications to Lyman and Balmer Series]{A Theory of Quantum Jumps
and its Applications to Lyman and Balmer Series}


\author*[1]{\fnm{Z. E.} \sur{Musielak}}\email{zmusielak@uta.edu}

\affil*[1]{\orgdiv{Department of Physics}, \orgname{University of Texas at Arlington}, 
\orgaddress{\street{Science Hall}, \city{Arlington}, \postcode{76019}, \state{Texas}, \country{USA}}}


\abstract{A theory of quantum jumps is developed by using a 
new asymmetric equation, which is complementary to the 
Schr\"odinger equation.  The new equation is a mathematical
form of Bohr's rule for quantum jumps, and its solutions 
demonstrate that once a quantum jump takes place at a random 
time, then its evolution is continuous and coherent.  The temporal 
solutions are used to determine time-scales of the quantum jumps, 
which are found to be not instantaneous but finite.  The spatial 
solutions are used to obtain the radial probability density for an 
electron after its jump.  The obtained results are applied to the 
Lyman and Balmer series, and their experimental verification is 
discussed.
}




\maketitle

% Abstract (Do not insert blank lines, i.e. \\) 
\abstract{A theory of quantum jumps is developed by using a 
new asymmetric equation, which is complementary to the 
Schr\"odinger equation.  The new equation is a mathematical
form of Bohr's rule for quantum jumps, and its solutions 
demonstrate that once a quantum jump takes place at a random 
time, then its evolution is continuous and coherent.  The temporal 
solutions are used to determine time-scales of the quantum jumps, 
which are found to be not instantaneous but finite.  The spatial 
solutions are used to obtain the radial probability density of an 
electron after its jump.  The obtained results are applied to the 
Lyman and Balmer series, and their experimental verification is 
discussed.
}

\section{Introduction}
  
Quantum jumps are transitions of electrons between discrete 
energy levels in atoms when electromagnetic radiation of a specific 
energy is absorbed or emitted by the atoms.  In the original hydrogen 
atom model [1], Bohr postulated that the passing between different 
quantum states in the atom, when electrons change their allowed 
energy levels by absorbing or emitting photons of very specific 
energies, cannot be properly described by mechanics.  Einstein [2]
introduced the notion that such transtions are random and questioned
their randomness, while Bohr responded by arguing that the transitions 
are instantaneous.  As pointed out by Schr\"odinger [3], later the 
transitionswere named 'quantum jumps' or were also called Bohr-Einstein 
quantum jumps.  Schr\'odinger [3] argued against the jumps being 
instantaneous as no process in the real world happens instantenously 
or in zero time.  Then, the idea of quantum jumps being random and 
instantaneous, which has been reapetedly verified experimentally, 
has become associated with the Copenhagen interpretation of 
quantum mechanics (e.g., [2-5]).

The first direct observations of quantum jumps in individual 
atoms exposed to electromagnetic (EM) fields took place in 
1986 and were reported in [7-9].  The collected data showed 
that the atoms flipped between a state that allowed them to 
emit a photon to a state that did not, and that the atoms 
remain in one state, or the other, for periods ranging from 
a few tenths of a second to a few seconds before jumping 
again.  Other independent experiments confirmed the results 
and demonstrated that the jumps were random and abrupt 
(e.g., [10-12], and references therein).  A different view has 
emerged from the experimental results obtained by Minev 
et al. [13] confirming that quantum jumps occur at random 
times, but once they occur, the evolution of each completed 
jump appears to be a continuous and coherent physical 
process that unfolds in a finite time.  

There were many attempts to formulate theories of quantum 
jumps in atomic systems aiming to relate the resulting theoretical 
predictions to the experimental data available at the time when 
such theories were developed (e.g., [14-19]).  Specifically, it was 
suggested that the so-called quantum trajectory theory (QTT) 
may account for the experimental data presented in [13].  QTT 
gives trajectories of individual particles that obey the probabilities 
computed from the Schr\"odinger equation [20,21], and the theory 
is used to study both open and dissipative quantum systems. 
There are three different quantum trajectories considered in QTT:
non-real quantum trajectories treated as numerical tools 
[20], and subjectively [21] and objectively [22] real quantum 
trajectories (e.g., [23]).  A trajectory in QTT is not like a 
classical trajectory, but instead it is a path in Hilbert space that 
is determined by solving the stochastic Schr\"odinger equation 
(e.g., [23]).  Predictions made by QTT cannot be reproduced
by the standard quantum mechanics approach based on the 
Schr\"odinger equation (e.g., [24]). 

More recently, a theory of quantum jumps was developed for 
atoms that are coupled to a quantized electromagnetic field in the  
limiting regime where the orbital motion of the atoms is neglected
and the velocity of light is assumed to be infinitely large [25].  
It must also be pointed out that quantum jumps play important 
roles in quantum feedback control [26,27], and in detecting and 
fixing decoherence-induced errors in quantum information systems 
[28,29].

The theoretical approach presented in this paper is very different 
from any previous attempt to account for quantum jumps.  The 
new theory of quantum jumps developed herein is based on a 
new asymmetric equation (NAE) [30] that was derived from the 
irreducible representations (irreps) of the Galilean group of the 
metrics (e.g., [31-33]).  Since the same irreps were used 
to derive the Schr\"odinger equation [34,35], the equations are 
considered to be complementary to each other [30].  The main 
physical implication of this complementarity is that some quantum 
phenomena, like the quantum jumps considered in this paper, or 
the quantum measurement problem [36], cannot be accounted for 
by the Schr\"odinger equation and they require the NAE to describe 
them. 

The NAE naturally describes interactions of quantum systems 
with environments that are subjected to monitoring by measuring 
devices.  Specifically, the NAE can be used to describe the dynamics 
of quantum particles during their transitions in atoms as shown in 
this paper, or interactions of a single quantum particle with a 
measuring apparatus, as demonstrated in [36], or other physical 
processes [30].  The main reason is that non-unitary theories of
these quantum processes can be developed using the NAE because 
of its second-order derivatives in time (see Section 2, for details).  
Random and instantaneous quantum jumps violate the unitary 
evolution, and they cannot be represented by the Schr\"odinger 
equation (e.g., [2-13]); thus, in this paper, a theory of quantum 
jumps is developed based on the NAE.

The main aim of this paper is to develop a theory of  quantum 
jumps using the NAE.  The theory is applied to a hydrogen 
atom and used to describe absorption Lyman and Balmer 
series.  It is shown that Bohr's rule for quantum jumps 
(e.g., [3-5]) emerges naturally from the NAE, and that the 
NAE can be solved analytically by separation of variables.  
To preserve the coherence of the wavefunction, the solutions
are restricted to the EM frequencies corresponding directly to 
the jumps; in other words, measurements that destroy the 
coherence are not accounted for by the solutions.  The 
temporal solutions of the NAE show that time-scales for 
the quantum jumps are very short but finite.  The spatial 
solutions are used to obtain the radial probability densities, 
which are normal distributions with the location of their 
centers in agreement with Bohr's rule for quantum jumps.  
Experimental verification of the obtained results is also 
discussed.

The paper is organized as follows: the basic equations resulting 
from the Galilean group of the metrics are given and discussed 
in Section 2; the governing equation and its solutions are 
presented in Section 3;  the solutions are applied to the Lyman 
and Balmer absorption processes in Section 4;  experimental 
verification of the obtained results is discussed in Section 5; 
and Conclusions are given in Section 6. 

\section{Schr\"odinger and new asymmetric equations}

In Galilean relativity, space and time are separated and their 
metrics are given by $ds_1^2\ =\ dx^2 + dy^2 + dz^2$ and 
$ds_2^2\ =\ dt^2$, where $x$, $y$ and $z$ are spatial 
coordinates and $t$ is time.  The transformations that make 
the metrics invariant form the Galilean group of the metric,
${\mathcal G}$, whose structure is ${\mathcal G} \ = S (4) 
\otimes_s H (6)$, where $S(4) = T(1) \otimes R(3)$ is a 
four-parameter subgroup of translation in time and rotations 
in space, and $H(6) = T(3) \otimes B(3)$ a six-parameter 
subgroup translations in space and boosts [32,33].  

The group ${\mathcal G}$ is a ten-parameter Lie group and $H(6)$ 
is its invariant Abelian subgroup.   It is well-known that the irreducible 
representations (irreps) of $H(6)$ are one-dimensional [29-31], and 
they provide labels for all the irreps of ${\mathcal G}$.  Classification 
of the irreps of this group by Bargmann [29] demonstrated that the 
scalar and spinor irreps are physical, but the vector and tensor irreps 
are not because they do not allow for the elementary particle localization.  
Therefore, in nonrelativistic quantum mechanics (NRQM) only scalar and 
spinor wavefunctions are allowed and their temporal and spatial evolution 
is described by the Schr\"odinger [2-5] and L\'evy-Leblond [32,33] 
equations, respectively.  In this paper, only the scalar irreps are 
considered.  

Writing separately the transformations for space translations and boosts, 
one finds: $\hat {T}_{\mathbf {a}} \phi(\mathbf {r}, t)\ \equiv\ \phi
(\mathbf {r} + \mathbf {a}, t)\ = e^{i \mathbf {k} \cdot \mathbf {a}} 
\phi (\mathbf {r}, t)$ and $\hat {B}_{\mathbf {v}} \phi(\mathbf {r}, t)\ 
\equiv\ \phi(\mathbf {r} + \mathbf {v}, t)\ =\ e^{i \mathbf {k} \cdot 
\mathbf {v} t} \phi(\mathbf {r}, t)$, where $\phi(\mathbf {r}, t)$ is a 
scalar wavefunction, $\mathbf {a}$ represents a translation in space, 
$\mathbf {v}$ is the velocity of Galilean boosts, and the real vector 
$\mathbf {k}$ labels the one-dimensional irreps of $H(6)$.  Expanding 
$\phi(\mathbf {r} + \mathbf {a}, t)$ and $\phi(\mathbf {r} + \mathbf 
{v}t, t)$ in Taylor series and comparing the results to the above 
transformations, one obtains the following eigenvalue equation [34-36]
\begin{equation}
- i \nabla \phi(\mathbf {}r, t)\ =\ \mathbf {k} \phi (\mathbf {}r, t)\ ,
\label{eq1a}
\end{equation}

\noindent
which has the same form for the boosts and translations in space, and it 
is the necessary condition that the wavefunction $\phi (\mathbf {}r, t)$ 
transforms as one of the irreps of ${\mathcal G}$.

Now, $S(4)$ is not an invariant subgroup of ${\mathcal G}$, thus, the 
above procedure cannot be applied to it; the fact that $T(1)$ is an invariant 
subgroup of $S(4)$ does not help because $T(1)$ is not the 'little group' 
of ${\mathcal G}$ [31,34].  In the solution proposed in [34], a new function, 
$\phi_{\omega}(\mathbf {r}, t) = \eta(\mathbf {r}, t) \phi(\mathbf {r}, t)$, 
where $\eta(\mathbf {r}, t)$ is a smooth and differentiable function to be 
determined, and the generator of translation, $\hat E = i \partial_t = i 
\partial / \partial t$, were introduced, with $i \partial_t \phi_{\omega}
(\mathbf {r}, t)\ =\ \omega \phi_{\omega}(\mathbf {r}, t)$.  Then, it 
was demonstrated [34] that the Galilean Principle of Relativity requires 
$\eta (\mathbf {r}, t) = \eta^{\prime}(\mathbf {r}^{\prime}, t^{\prime}) 
= 1$, which means that 
\begin{equation}
i \partial_t \phi (\mathbf {r}, t)\ =\ \omega \phi 
(\mathbf {r}, t)\ ,
\label{eq1b}
\end{equation}
and that this eigenvalue equation supplements Eq. (\ref{eq1a}).
The eigenvalue equations given by Eqs (\ref{eq1a}) and (\ref{eq1b})
guarantee that the wavefunction $\phi (\mathbf {r}, t)$ transforms
as one of the irreps of ${\mathcal G}$, and that these equations 
can be used to derive dynamical equations that are consistent 
with Galilean Relativity.  

It must be noted that a customary approach of finding a dynamical 
equation from the Casimir operator of ${\mathcal G}$ cannot be 
used here because the operator does not connect Galilean space 
and time.  Thus, requires that ${\mathcal G}$ is modified to allow
 for such connections, which gives the extended Galilean group, 
whose structure is $\mathcal{G}_e\ = [R(1) \otimes B(3)] \otimes_s 
[T(3+1) \otimes U(1)]$, where $U(1)$ is a one-parameter unitary 
group (e.g., [32,33]); the structure of this group is similar to the 
Poincar\'e group (e.g., [37-39]).  There are three Casimir operators 
of $\mathcal{G}_e$, but for a scalar wavefunction only one of them 
gives the Schr\"odinger equation.  However, the problem with this 
approach is that $\mathcal{G}_e$ is not the group of the Galilean 
metrics, thus, the derivation of dynamical equations using the 
Casimir operator of $\mathcal{G}_e$ may not be consistent 
with Galilean Relativity. 

In the following, the eigenvalue equations, which are consistent 
with both the irreps of $\mathcal{G}$ and Galilean Relativity,  
together with the de Broglie relationship (e.g., [3-5]), are used 
to obtain the following two asymmetric equations [36]:
\begin{equation}
\left [ i {{\partial} \over {\partial t}} +  \frac{\hbar}{2 m}
\nabla^{2} \right ] \phi_S (t, \mathbf {x}) = 0\ ,
\label{eq2a}
\end{equation}  
which is the Schr\"odinger equation (SE) (e.g., [3-5]), and 
\begin{equation}
\left [ \frac{i}{\omega} {{\partial^2} \over {\partial t^2}} 
+ \frac{\hbar}{2 m}{\mathbf k} \cdot \nabla \right ] \phi_A 
(t, \mathbf {x}) = 0\ ,
\label{eq2b}
\end{equation}  
which is called the new asymmetric equation (NAE) [30]. 
Since both equations are derived from the same irreps,
they are considered to be complementary to each other.
The equations describe free particles, and they must be 
quantum particles because of the presence of the Planck 
constant appears as a coefficient in both equations.  

The SE is used to compute the probability of finding a 
quantum system in its possible states before any interaction 
of the system with its environment, or any measurements 
are performed on it.  The SE is useful for predicting average 
measurements of large ensembles of quantum objects.  The 
NAE describes and provides insight into the behaviour of 
individual particles affected by their surroundings represented 
by $\omega$ and $\mathbf {k}$ in Eq. (\ref{eq2b}); this 
treatment of individual particles shows some similarities 
between QTT and the theory presented in this paper.

Another important difference between the SE and NAE 
is that the former is first-order in its time derivative, thus 
allowing for unitary theories, while the latter can also be 
used to construct non-unitary theories because of its 
second-order time derivative.  In addition, the NAE 
requires that the eigenvalues $\omega$ and $\mathbf {k}$,
which are explicitly present in the equation, are specified;
this means that the NAE naturally allows to include quantum 
system's interaction with its environment.  In other words, 
by specifying these eigenvalues, an inertial observer 
becomes an active Galilean observer who changes 
the system (e.g., performs a measurement) and uses 
the NAE to describe the resulting changes, which cannot 
be done by a passive Galilean observer who uses the SE. 

There are processes in nonrelativistic QM, such as the 
quantum measurement problem and quantum jumps 
that are considered to be non-unitary [3-5].  In these 
processes, the external input/output of EM radiation 
plays an important role, and the NAE directly allows 
accounting for this radiation by specifying its 
$\omega$ and $\mathbf {k}$.  In this paper, the 
NAE is used to formulate a theory of quantum jumps.

\section{Model of quantum jumps}

\subsection{Formulation}

To model quantum jumps, a single hydrogen atom with its 
electron in the ground state is considered.  Quantum jumps 
are casual because they are triggered by EM radiation, and 
the NAE given by Eq. (\ref{eq2b}) is used to describe them.
The NAE predicts a new state of the electron after it jumps 
from the initial energy level, which is one of the s-orbitals, 
and it gives the duration time for the jump, and also finding 
the radial probability density of the electron at its new state.  
Now, for the electron’s final state, it is assumed that the 
electron returns to its original state either immediately 
after the jump, or after some time that can be established 
experimentally.  

The considered model of quantum jumps is based on the 
NAE with the Coulomb potential included in it.  Moreover, 
some results originally obtained from the SE with the 
Coulomb potential are also used in solving the NAE; 
specifically, the hydrogen atom energy levels, and the 
frequencies and wavevectors of EM radiation that are 
used to trigger quantum jumps.  As shown in the previous 
section, the SE and NAE are complementary, and and both 
equations contribute to the model.  However, the main 
difference is that the SE describes quantum unitary 
processes, while the NAE with its second-order time 
derivative accounts properly for non-unitary processes, 
such as quantum jumps.

\subsection{The governing equation}

To apply the NAE to quantum jumps, it is required that 
the labels of the irreps $\omega$ and $\mathbf {k}$ are 
specified.  Let $\omega = \omega _{o}$, $E_{o} = \hbar 
\omega_{o}$, and ${\mathbf {k}} = {\mathbf {k _{o}}}$, 
where $\omega_{o}$ and ${\mathbf {k _{o}}}$ represent 
frequency and wavevector of EM radiation absorbed or 
emitted by a hydrogen atom.  Then, Eq. (\ref{eq2b}) can 
be written in the following form 
\begin{equation}
\left [ \frac{i \hbar}{\omega _{o}} \left ( {{\partial^2} \over 
{\partial t^2}} \right ) + \frac{\hbar^2}{2 m} ( {\mathbf k _{o}} 
\cdot \nabla ) + V(r) \right ] \phi_A (t, \mathbf {r}) = 0\ ,
\label{eq3}
\end{equation}  
where $V(r) = - e^2 / (4 \pi \epsilon r)$ is the Coulomb potential,
which is included into the NAE to accout properly for the energy 
levels in the hydrogen atom.  Since the SE is used to compute 
the probability of finding a quantum system in its possible state
before any interaction with the system's environment or any 
measurement performed on the system, the NAE can be used 
to describe the system that interacts with its environment by 
specifying $\omega_{o}$ and ${\mathbf {k _{o}}}$; note 
that if $\omega_{o} = 0$ and ${\mathbf {k _{o}}= 0}$, 
then the system must be described by the SE.  This shows 
the complementary role these two equations play in NRQM. 

The parameters $\omega_{o}$ and ${\mathbf {k _{o}}}$ in 
the above equation are determined by using the solutions to the 
time-independent SE, which provide the wavefunctions and the 
energy levels associated with these wavefunctions.  The energy 
levels are given by
\begin{equation}
\Delta E = R_{E} \left ( \frac{n_f^2 - n_i^2}{n_f^2 n_i^2}
\right )\ ,
\label{eq4}
\end{equation}  
where $R_{E} = \hbar^2 / 2m a_B^2$ is the Rydberg constant, 
$a_B = (4 \pi \epsilon_o \hbar^2) / m e^2$ is the Bohr radius,
and $n_i$ and $n_f$ are the prinicipal quantum numbers 
corresponding to the initial and final states, respectively. 
For absorption, $n_f > n_i$ and $\Delta E > 0$; however,
for emission, $n_f < n_i$ and $\Delta E < 0$.  Taking 
$\vert \Delta E \vert = \hbar \omega_o$, the parameters 
$\omega_{o}$ and ${\mathbf {k _{o}}}$ are determined 
from Eq. (\ref{eq4}) that gives
\begin{equation}
\omega_o = \frac{R_{E}}{{n_i^2 n_f^2} \hbar}  \vert  n_f^2 
- n_i^2 \vert\ ,
\label{eq5}
\end{equation}  
{\bf and} 
\begin{equation}
{\mathbf {k_o}} = k_o {\mathbf {\hat {k}}_{o}} = \frac{\omega_o}
{c} {\mathbf {\hat {k}}_{o}} = \frac{R_{E}}{{n_i^2 n_f^2} c \hbar} 
\vert n_f^2 - n_i^2 \vert {\mathbf {\hat {k}}_{o}}\ ,
\label{eq6}
\end{equation}  
which guarantee that $\omega_o$ and $\mathbf {k_o}$ are positive 
for both absorption and emission.

Then, Eq. (\ref{eq3}) becomes 
\begin{equation}
\left [ \frac{i \hbar}{\omega _{o}} \left ( {{\partial^2} \over 
{\partial t^2}} \right ) + \left ( \frac{\lambda_C}{4 \pi} \right )
(\hbar \omega_o) ({\mathbf{\hat {k}}_{o}} \cdot \nabla ) - 
\left ( \frac{\hbar^2}{m a_B} \right ) \frac{1}{r} \right ] 
\phi_A (t, \mathbf {r}) = 0\ ,
\label{eq7}
\end{equation}  
which is  the governing equation for the presented theory, with 
$\lambda_C = h / mc$ denoting the Compton wavelength.

\subsection{Temporal and spatial solutions}

To solve the governing equation, the spherical variables are separated 
into the temporal and spatial (radial only) components, $\phi_A (t, 
\mathbf {r}) = \chi (t)\ \eta (\mathbf {r})$, and the separation 
constant $- \mu^2  = E_n = - (\hbar^2 / 2m) (1 / n a_B)^2$.  
Then,  Eq. (\ref{eq7}) becomes
\begin{equation}
{{d^2 \chi} \over {d t^2}} + i \left  ( \frac{\hbar \omega_o}
{m n^2 a_B^2} \right ) \chi = 0\ ,
\label{eq8}
\end{equation}  
and
\begin{equation}
{{d \eta} \over {d r}} + \frac{2}{(\mathbf {\hat {k} _{o}} 
\cdot \mathbf {\hat {r}}) k_{o} a_B} \left  ( \frac{1}{n^2 
a_B} - \frac{1}{r} \right ) \eta = 0\ ,
\label{eq9}
\end{equation}  
where $n$ is the principal quantum number; it is seen that the 
time-independent NAE directly displays Bohr's rule.  Since the
solutions of both equations depend on $n$, let $\chi (t) = \chi_n (t)$ 
and $\eta (r) = \eta_n (r)$.  Note that $n = n_f$ for absorption, 
and $n = n_i$ for emission.

Taking $i = (1/ \sqrt{2} + i / \sqrt{2})^2$, Eq. (\ref{eq8}) 
can be integrated giving solutions  
\begin{equation}
\chi_n (t) = C_{\pm} \exp{ \left [ \pm\ i \left ( \frac{1}{\sqrt{2}} 
+ \frac{i}{\sqrt{2}} \right ) \Omega_n t \right ]}\ ,
\label{eq10}
\end{equation}  
where $C_{\pm}$ are the integration constants corresponding 
to the $\pm$ solutions, and the characteristic frequency
$\Omega_n$ is given by   
\begin{equation}
\Omega_n = \frac{1}{n a_B} \sqrt{\frac{\hbar \omega_o}{m}}\ .
\label{eq11}
\end{equation}  
Both solutions with $C_{+}$ and $C_{-}$ are physical and they 
correspond to $t \rightarrow + \infty$ and $t \rightarrow - \infty$,
respectively.  In the following, only the solution with $C_{+}$ is 
considered because quantum jumps occur when $t > 0$.  The 
real part of the solution is 
\begin{equation}
{\cal {R} \it e} [\chi_n (t)] = C_{+} \cos{( \Omega_n t)}  \exp{( - 
\Omega_n t)}\ ,
\label{eq12}
\end{equation}  
and its physical meaning and applications to a hydrogen atom are
presented and discussed in Setion 4. 

Before solving Eq. (\ref{eq9}), it must be noted that in spherical 
symmetry $\mathbf {\hat r}$ can always be aligned with $\mathbf 
{\hat {k} _{o}}$ for the absorbed EM radiation, which means that 
$\mathbf {\hat {k} _{o}} \cdot \mathbf {\hat {r}} = 1$.  Similarly,
for EM radiation emitted by an atom, the direction of such radiation 
is undetermined until it is measured by macroscopic instruments; 
therefore, $\mathbf {\hat {k} _{o}} \cdot \mathbf {\hat {r}} = 1$
remains also valid for emission.  

It is also convenient to transform Eq. (\ref{eq9}) to the new independent 
variable $r_a = r / a_B$, which gives
\begin{equation}
\frac{d \eta_n}{d r_a} + \beta_B \left  ( \frac{1}{n^2} - \frac{1}
{r_a} \right ) \eta_n = 0\ ,
\label{eq13}
\end{equation}  
where 
\begin{equation}
\beta_B = 8 \pi \left ( \frac{a_B}{\lambda_c} \right ) \left ( \frac{n_f^2 
n_i^2}{n_f^2 - n_i^2} \right )\ .
\label{eq14}
\end{equation}  
According to Eq. (\ref{eq13}),  $\eta_n (r)$ reaches its maximum value 
when $r_a = n^2$ or $r = n^2 a_B$, which is Bohr's rule for quantum 
jumps (e.g., [3-5]).  In other words, Eq. (\ref{eq13}) is the mathematical 
form of Bohr's rule, and it gives the most probable radius $r_a$, where 
the maximum value of $\eta_n (r)$ occurs.

Having obtained the time-independent NAE given by Eq. (\ref{eq13}),
it is seen that its form is different than a typical time-independent SE 
used in applications in NRQM (e.g., [4,5]).  By being a second-order 
ODE, the SE can be considered as a Sturm-Liouville problem with its 
resulting eigenvalue equation; however, the NAE being a first-order 
ODE cannot be cast in the same form.  Instead, the NAE requires 
specifying its eigenvalues $\omega_o$ and $\mathbf {{k} _{o}}$ 
as described above.   Moreover, Eq. (\ref{eq13}) directly displays 
Bohr's rule, and its shows that the the time-independent wavefunction 
$\eta_n (r)$ reaches its minimum at $r = n^2 a_B$.  Bohr's rule can 
also be obtained from the time-independent SE through the construction 
of hydrogen wave functions that satisfy the following condition $<1/r> 
= 1/(n^2 a_B)$ (e.g., [4,5]).

The solution to Eq. (\ref{eq13}) is 
\begin{equation}
\eta_n (r) = \eta_{o,n}\ r_a^{\beta _{B}}\ \exp{ \left [ - \left ( 
\frac{\beta _{B}}{n^2} \right ) r_a\right ]}\ ,
\label{eq15}
\end{equation}  
where the integration constant $\eta_{o,n}$ is obtained from 
the following normalization condition  
\begin{equation}
4 \pi a_B^3 \int_0^{\infty} r_a^2 \vert \eta_n (r_a) \vert^2 dr_a = 1\ .
\label{eq16}
\end{equation}  
For $\eta_n (r)$ given by Eq. (\ref{eq15}), the integral can be evaluated
and the result is 
\begin{equation}
\eta_{o,n}^2 (\beta_B) = \frac{1}{4 \pi a_B^3 \Gamma (2 \beta_B + 3)}
\left ( \frac{2 \beta_B}{n^2} \right )^{2 \beta_B + 3}\ ,
\label{eq17}
\end{equation}  
which is valid if ${\cal {R} \it e} (2 \beta_{B}) > 0$ and 
${\cal {R} \it e} (4 \beta_{B} + 2) > -1$; note that both 
conditions are obeyed in the theory presented herein.

\subsection{Radial probability density}

After finding the normalized spatial wavefunction $\eta_n (r)$, 
the radial probability density in a spherical shell volume element 
can be evaluated, and it is given by  
\begin{equation}
dP_n (r_a) = 4 \pi a_B^3 r_a^2 \vert \eta_n (r_a) \vert^2 dr_a\ ,
\label{eq18}
\end{equation}  
or, after using Eqs (\ref{eq15}) and (\ref{eq17}), one obtains
\begin{equation}
{\cal{P}}_{n} (r_a) \equiv \frac{dP_n (r_a)}{d r_a}  = \frac{1}
{\Gamma (2 \beta_B + 3)} \left (\frac{2 \beta_B}{n^2}
\right )^{2 \beta_{B} + 3}\ r_a^{2 ( \beta_{B} + 1)}\ 
e^{-(2 \beta_{B} / n^2) r_a}\ ,
\label{eq19}
\end{equation}  
which gives $\int_0^{\infty} {\cal{P}}_{n} (r_a) d r_a = 1$.
The obtained ${\cal{P}}_{n} (r_a)$ gives the radial probability 
density for the electron being in this state after the transition.

\subsection{Physical meaning of the solutions}

Having obtained the temporal and spatial solutions to the 
NAE, the physical meaning of these solutions is now discussed.  
According to Eq. (\ref{eq12}), the temporal component $\chi_n (t)$ 
of the wavefunction $\phi_A (t, \mathbf {r})$ decays exponentialy 
in time, and the rate of this exponential decay depends on the 
value of the characteristic frequency $\Omega_n$ given by Eq. 
(\ref{eq11}).  Let $T_n = 2 \pi / \Omega_n$ be the duration 
time of a single quantum jump, then, the results presented 
below show that this time is very short but finite.  By calculating 
this time for a considered quantum jump, and knowing the distance 
travelled by the electron during the jump, the electron speed can
be calculated for each jump.

The spatial solution and its normalization factor given by Eqs 
(\ref{eq15}) and (\ref{eq17}), respectively, were used to obtain 
the radial probability density for any quantum jump (see Eq. 
\ref{eq19}).  The theory predicts that after an electron makes 
a quantum jump, the probablility density of finding it is very 
narrow and centered at $r = n^2 a_B$, which is consistent 
with Bohr's rule for quantum jumps.  Experimental results 
[6-8] demonstrated that the electron remains in the excited 
state for periods ranging from a few tenths of a second to a 
few seconds before jumping again, and returns to its original 
state.  The developed theory describes such physical situations, 
and its solutions preserve the coherence of the wavefunction.
To achieve it, the solutions are restricted only to the EM  
frequencies corresponding directly to the quantum jumps 
(see Sec. 4).  In other words, the solutions do not account 
for measurements because they destroy the coherence of 
the wavefunction; to retain this coherence, Minev et al. 
[13] employed a special procedure in their experiment.

In the following, the obtained theoretical results are applied
to a hydrogen atom and its Lyman and Balmer series absorption.
The theory also describes emission of EM radiation when the 
initial, $n_i$, and final, $n_f$, are properly defined for such
transitions.

\section{Application to hydrogen atom}

\subsection{Lyman series absorption}

The developed theory is now applied to a hydrogen atom 
by considering absorption of EM radiation that causes the 
electron's transitions from its initial, $n_i = 1$, to its final,
$n_f = 2$, $3$ or $4$ energy levels.  For this Lyman series 
absorption, the frequencies $\omega_o$ and $\Omega_n$ 
are calculated from Eqs (\ref{eq5}) and (\ref{eq11}), which 
give 
\begin{equation}
\omega_o = 2.065 \cdot 10^{16} \left ( \frac{n_f^2 - n_i^2}
{n_f^2 n_i^2} \right )\ {\rm [ s^{-1}]}\ ,
\label{eq20}
\end{equation}  
and 
\begin{equation}
\Omega_n = 2.03 \cdot 10^{8}\ \frac{\sqrt{\omega_o}}{n}\
{\rm [ s^{-1}]}\ ,
\label{eq21}
\end{equation}  
where $n = n_f$.  The characteristic frequency $\Omega_n$ can 
be used to estimate the duration time of quantum jumps $T_n = 
2 \pi / \Omega_n$.  Knowing the distance traveled by the electron 
and the duration time, the electron's speed $v_e = (n_f^2 - n_i^2) 
a_B / T_n$ can also be calculated.  
\smallskip
\begin{table*}[t]
\begin{center}
\centering 
\begin{tabular}{|c| c| c| c| c|c|} 
\hline 
\hline 
Lyman series & $n = n_f$ & $\omega_o$ [$s^{-1}$] & $\Omega_n$ 
[$s^{-1}$] & $T_n$ [s] & $v_e$ [$m\ s^{-1}$]\\ [0.5ex] 
\hline
Ly-$\alpha$ & 2 & $1.549 \cdot 10^{16}$ & $1.265 \cdot 10^{16}$ 
& $5.0 \cdot 10^{-16}$ & $3.2 \cdot 10^{5}$\\ 
Ly-$\beta$ & 3 & $1.836 \cdot 10^{16}$ & $9.171 \cdot 10^{15}$ 
& $6.9 \cdot 10^{-16}$ & $6.2 \cdot 10^{5}$\\ 
Ly-$\gamma$ & 4 & $1.936 \cdot 10^{16}$ & $7.061 \cdot 10^{15}$ 
& $8.9 \cdot 10^{-16}$ & $8.9 \cdot 10^{5}$\\ 

\hline 
\end{tabular}

\smallskip
Table 1. Theoretical predictions of the characteristic frequency $\Omega_n$,
the duration time $T_n$, and the speed of electron $v_e$ during quantum 
jumps corresponding to the Lyman series resulting from absorption of EM 
radiation with frequency $\omega_o$.\\
\end{center}
\end{table*}
\smallskip

The obtained results are presented in Table 1.  It shows that the characteristic 
frequency $\Omega_n$ of quantum jumps is of the order of $(10^{15} - 
10^{16})$ $s^{-1}$ for the considered transitions in the Lyman series 
absorption; the frequency decreases for the quantum jumps to higher values 
of $n_f$.  Then, the corresponding duration time $T_n = 2 \pi / \Omega_n$ 
is of the order of $10^{-16}$ $s$, which is very short but yet finite.  

Since the distance the electron travels during one specific jump is known, 
and since the duration time of this jump is also estimated, the electron's 
speed is calculated.  For the three transitions considered for the Lyman 
series absorption, the speed is of the order of $10^5$ $m/s$ as shown
in Table 1; its value increases for the transition to higher orbitals.  

To calculate the radial probability density given by Eq. (\ref{eq19}), it is 
required to determine $\beta_B$, which is defined by Eq. (\ref{eq14})
and written as 
\begin{equation}
\beta_B = 548.2 \left ( \frac{n_f^2 n_i^2}{n_f^2 - n_i^2} \right )\ .
\label{eq22}
\end{equation}  
%
%

%
%%%%%%%%%%%%%%%%%%%%%%%%%%%%
\begin{figure}
\begin{center}
\includegraphics[width=115mm]{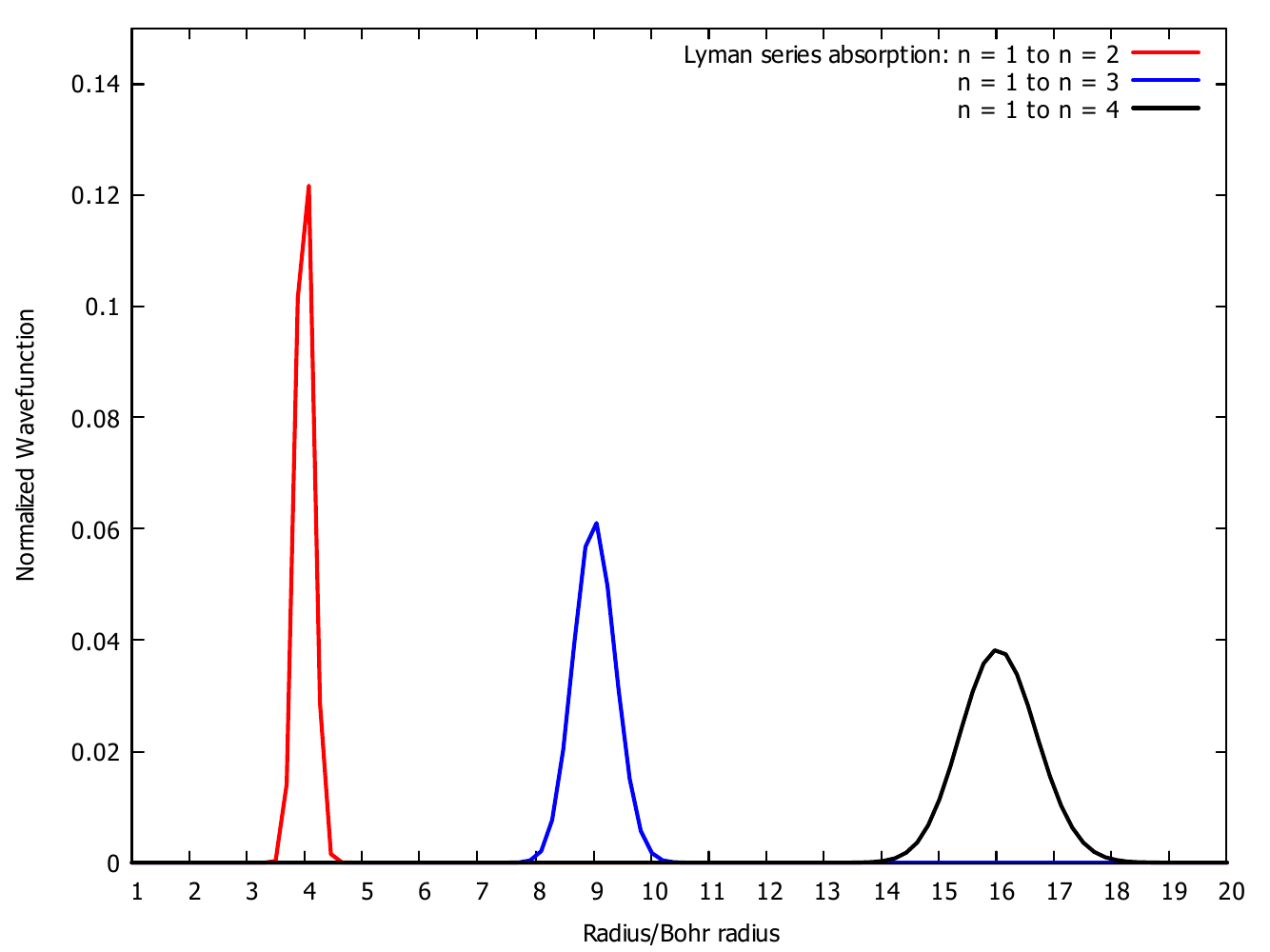}
\caption{The normalized wavefunction, $\eta_n (r_a)$, in 
units $a_B^{-3/2}$, is plotted versus the ratio of radius to 
the Bohr radius, $r / a_B$, for the Lyman series absorption 
from $n_i = 1$ to $n_f = 2$, $3$ and $4$.
}
\label{fig.1}
\end{center}
\end{figure}
%%%%%%%%%%%%%%%%%%%%%%%%%%%%
%

%
%%%%%%%%%%%%%%%%%%%%%%%%%%%%
\begin{figure}
\begin{center}
\includegraphics[width=115mm]{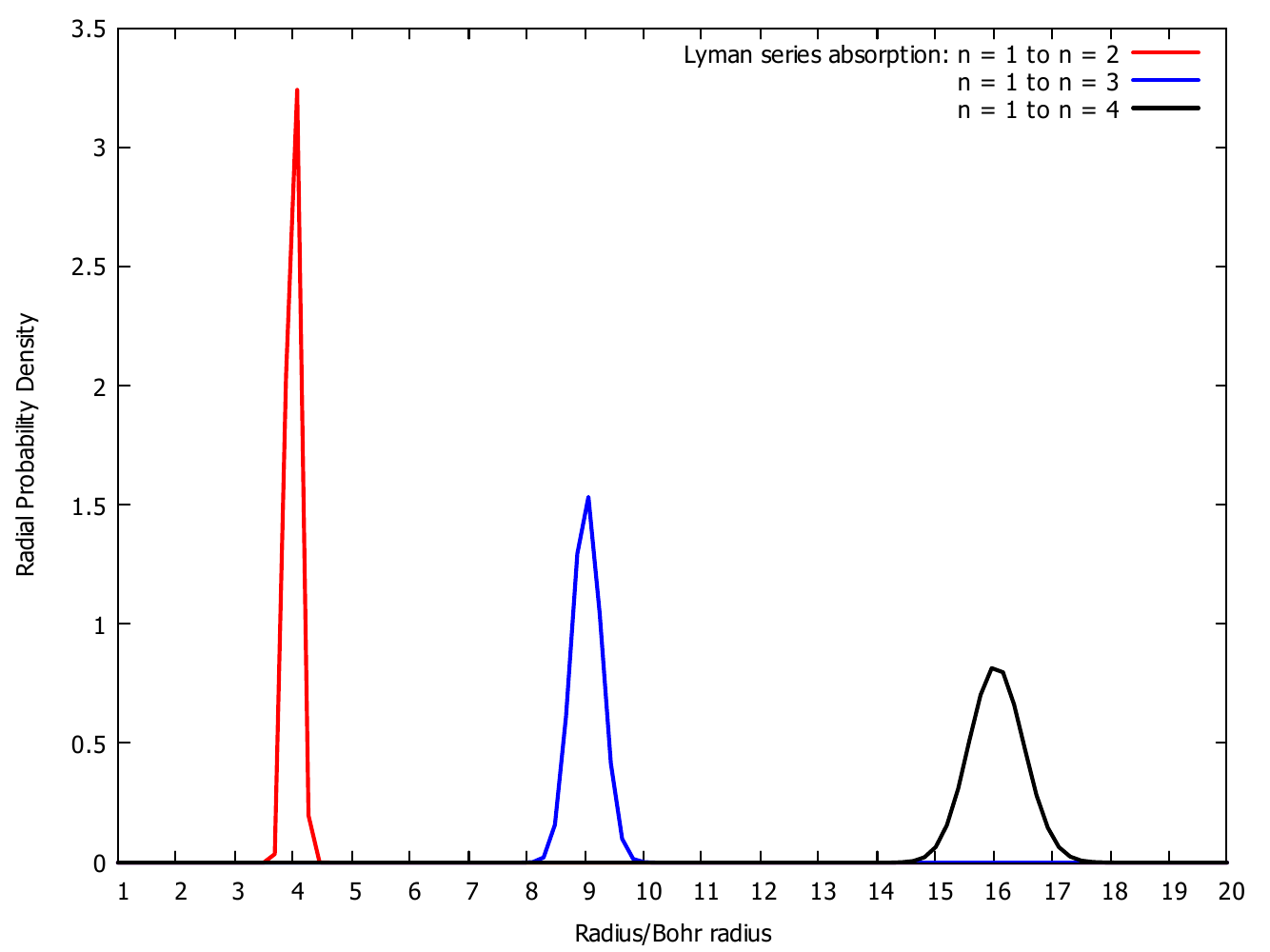}
\caption{The radial probability density ${\cal{P}}_{n} (r_a)$ 
in a spherical shell volume element corresponding to each 
transition is plotted versus the ratio of radius to the Bohr radius, 
$r / a_B$.  The plotted probability densities are for the Lyman 
series absorption from $n_i = 1$ to $n_f = 2$, $3$ and $4$.
}
\label{fig.2}
\end{center}
\end{figure}
%%%%%%%%%%%%%%%%%%%%%%%%%%%%
%

Having obtained $\beta_B$, the wavefunction $\eta_n (r_a)$ 
given by Eq. (\ref{eq15}) is calculated for the transitions 
from $n_1 = 1$ to $n_f = 2$, $n_f = 3$ and $n_f = 4$.  
The wavefunctions corresponding to these transitions are 
normalized (see Eq. \ref{eq17}) with its units being $a_B^{-3/2}$ 
(e.g., [4]), and plotted in Fig. 1.  The presented results demonstrate 
that the maxima of the wavefunctions are centered at $r_a$ = $4$, 
$9$ and $16$, respectively, which is in agreement with Bohr's rule 
for quantum jumps.  Moreover, the wavefunctions are symmetric 
about their corresponding maxima, and the shapes of the wavefunctions 
change from very narrow for the lowest $n_f$ to much wider for 
larger $n_f$. 

The computed wavefunctions are then used to calculate the radial 
probability density ${\cal{P}}_{n} (r_a)$ in a spherical shell volume 
element corresponding to each transition by using Eq. (\ref{eq19}).  
The obtained results are ploted in Fig. 2.  These are the probability 
densities for the electron being at its new location after the EM 
radiation required for the jump was absorbed.  The theory predicts 
that the resulting radial probablility densities are centered at $r_a$ 
= $4$, $9$ and $16$, respectively, as postulated by Bohr for quantum 
jumps (e.g., [4]).  The shapes of these probability curves range from 
very narrow for the lowest $n_f$ to much wider for higer values of $n_f$.  
Comparison of the results of Fig. 2 to those in Fig. 1 shows the effect 
of $r_a^2$ on the plotted ${\cal{P}}_{n} (r_a)$.  

The theory predicts the characteristic frequency ($\Omega_n \sim 
10^{16}$ $s^{-1}$), the time-scale ($T_n \sim 10^{-16} s$) 
for the considered jumps, which are used to estimate the electron's 
speed during the jumps; according to Table 1, the speeds 
are $v_e \sim 10^{5}$ $m/s$.  Moreover, the theory also 
gives the radial probability density of finding the electron in 
the excited state.  As the experiments [7-9] demonstrated, 
the electron spends from a few tenths of a second to a few 
seconds in the excited state before it jumps back to its 
original orbital.    

\subsection{Balmer series absorption}

The theory is now applied to the Balmer series absorption 
for the transitions from $n_i = 2$ to $n_f = 3$, $4$ and 
$5$, which correspond to the $H \alpha$, $H \beta$ and 
$H \gamma$ Balmer lines, respectively.   The obtained 
characteristic frequency, $\Omega_n$, the time-scale, 
$T_n$, and the electron's speed, $v_e$, are given in 
Table 2, which shows that $\Omega_n \sim 10^{15}$ 
$s^{-1}$, $T_n \sim 10^{-15}$ $s^{-1}$, and $v_e 
\sim 10^5$ $m/s$.  Comparison of these results to 
those obtained for the Lyman series absorption (see 
Table 1) indicates that while there is one order of 
magnitude difference in $\Omega_n$ and $T_n$, 
the electron's speed is of the same order for the all 
quantum jumps.
\smallskip
\begin{table*}[t]
\begin{center}
\centering 
\begin{tabular}{|c| c| c| c| c|c|} 
\hline 
\hline 
Balmer series & $n = n_f$ & $\omega_o$ [$s^{-1}$] & $\Omega_n$ 
[$s^{-1}$] & $T_n$ [s] & $v_e$ [$m\ s^{-1}$]\\ [0.5ex] 
\hline
H$\alpha$ & 3 & $2.868 \cdot 10^{15}$ & $3.623 \cdot 10^{15}$ 
& $1.7 \cdot 10^{-15}$ & $2.1 \cdot 10^{5}$\\ 
H$\beta$ & 4 & $3.872 \cdot 10^{15}$ & $3.158\cdot 10^{15}$ 
& $2.0 \cdot 10^{-15}$ & $3.2 \cdot 10^{5}$\\ 
H$\gamma$ & 5 & $4.336 \cdot 10^{15}$ & $2.673 \cdot 10^{15}$ 
& $2.4 \cdot 10^{-15}$ & $4.7 \cdot 10^{5}$\\ 
\hline 
\end{tabular}

\smallskip
Table 2. Theoretical predictions of the characteristic frequency $\Omega_n$,
the duration time $T_n$, and the speed of electrons $v_e$ during quantum 
jumps corresponding to the Balmer series resulting from absorption of EM 
radiation with frequency $\omega_o$.\\
\end{center}
\end{table*}
\smallskip
%

%
%%%%%%%%%%%%%%%%%%%%%%%%%%%%
\begin{figure}
\begin{center}
\includegraphics[width=115mm]{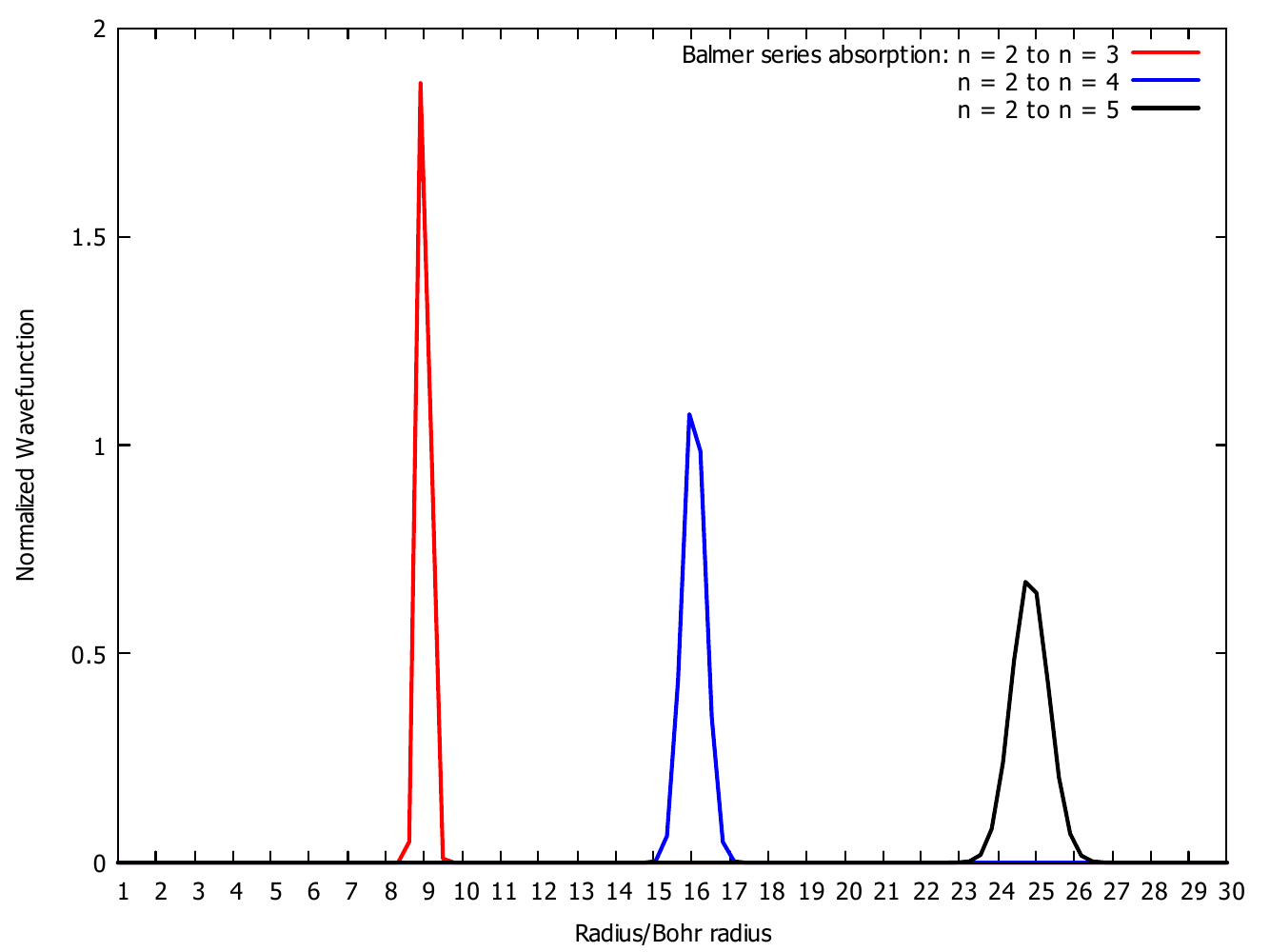}
\caption{The normalized wavefunction, $\eta_n (r_a)$,  in 
units $a_B^{-3/2}$, is plotted versus the ratio of radius to 
the Bohr radius, $r / a_B$, for the Balmer series absorption 
from $n_i = 2$ to $n_f = 3$, $4$ and $5$.
}
\label{fig.3}
\end{center}
\end{figure}
%%%%%%%%%%%%%%%%%%%%%%%%%%%%
%

%
%%%%%%%%%%%%%%%%%%%%%%%%%%%%
\begin{figure}
\begin{center}
\includegraphics[width=115mm]{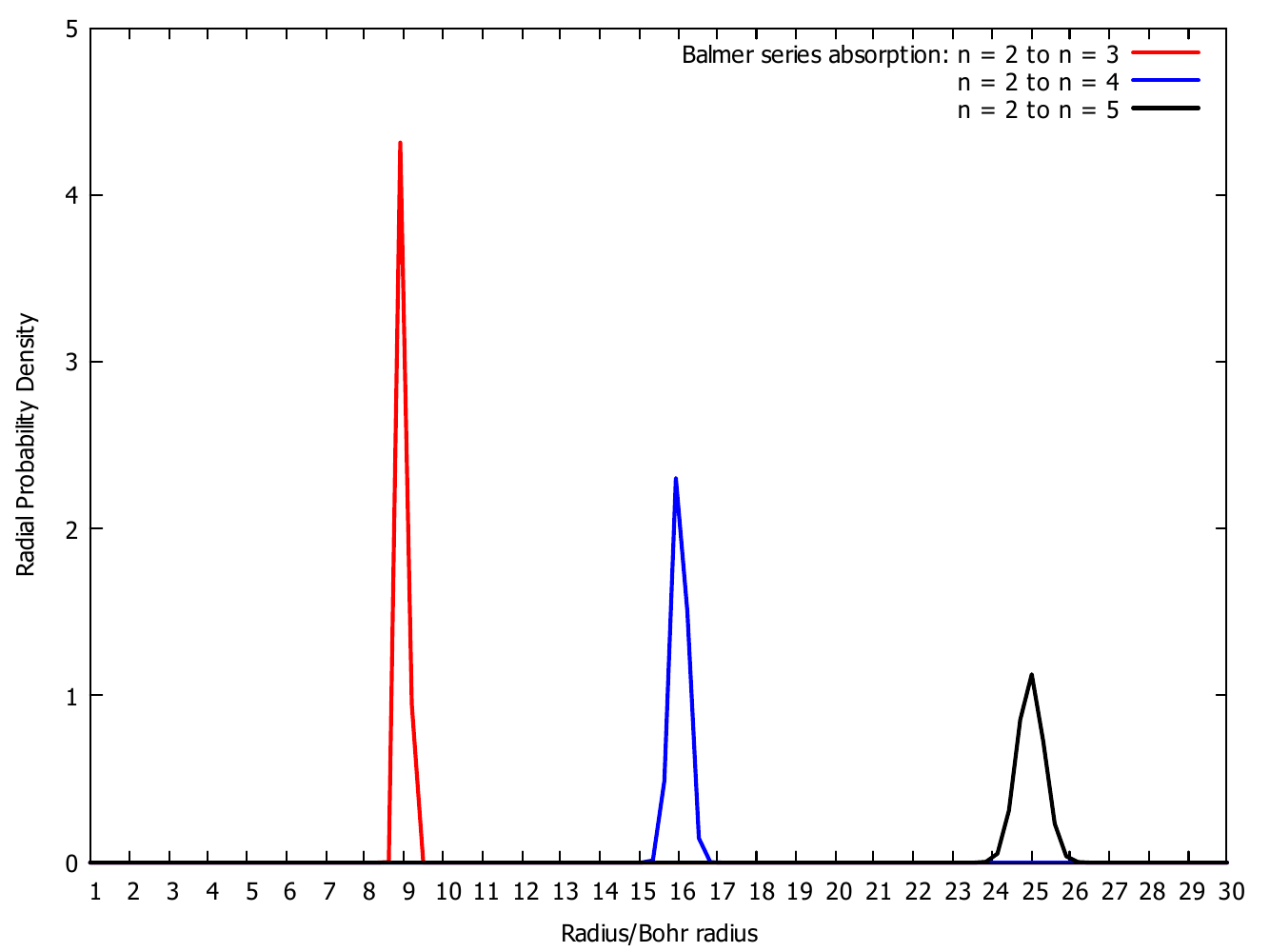}
\caption{The radial probability density ${\cal{P}}_{n} (r_a)$ 
in a spherical shell volume element corresponding to each 
transition is plotted versus the ratio of radius to the Bohr radius, 
$r / a_B$.  The plotted probability densities are for the Balmer
series absorption from $n_i = 2$ to $n_f = 3$, $4$ and $5$.
}
\label{fig.4}
\end{center}
\end{figure}
%%%%%%%%%%%%%%%%%%%%%%%%%%%%
%

The wavefunction $\eta_n (r_a)$ for the transitions from $n_1 = 2$ to 
$n_f = 3$, $n_f = 4$ and $n_f = 5$ is computed using Eq. (\ref{eq15}) 
and plotted in Fig. 3; the presented wavefunctions are normalized (see Eq. 
\ref{eq17}) and their units are $a_B^{-3/2}$ (e.g., [4]).  The location 
of the maximum of each plotted wavefuction is centered at $r_a$ = $9$, 
$16$ and $25$, respectively, which is consistent with Bohr's rule for 
quantum jumps.  Just as for the wavefunctions for the Lyman series 
absorption (see Fig. 1), the wavefunctions for the Balmer series 
absorption are also symmetric about their maxima, and their shapes 
change from very narrow for the lowest $n_f$ to much wider for 
larger $n_f$.  These may be the characteristic features of all the 
series absorptions in a hydrogen atom. 

Having obtained the normalized wavefunctions for the Balmer series 
absorption, Eq. (\ref{eq19}) is used to calculate the radial probability 
density ${\cal{P}}_{n} (r_a)$ in a spherical shell volume element 
corresponding to each transition, and the results are plotted in Fig. 4.  
As shown, the probability densities demonstrate that the most probable 
location of the electron after its absorption of the required EM radiation
is $r_a$ = $9$, $16$ and $25$, respectively, which is in agreement 
with Bohr's rule for quantum jumps (e.g., [4]).  The shapes of 
these probability curves change from very narrow for the lowest 
$n_f$ to much wider for higer values of $n_f$, similar as the 
probability densities for the Lyman series absorption shown in 
Fig. 2.  Moreover, comparing Figs 4 and 3, the effect of $r_a^2$ 
on the plotted ${\cal{P}}_{n} (r_a)$ is clearly seen.    

\subsection{Discussion of the obtained results}

The results presented for the Lyman and Balmer series absorption 
are obtained by using Eqs (\ref{eq5}) and (\ref{eq6}) to specify the 
values of $\omega_o$ and $\mathbf {k_o}$ .  Then, the NAE, which 
is a deterministic equation, is used to describe the transition from the 
initial state $n = n_i$  to the final state $n = n_f$.  The transition is 
causal because it is triggered by EM radiation with given $\omega_o$ 
and $\mathbf {k_o}$, and the theory predicts the physical characteristics 
of the resulting electron's jump.  

In the developed theory, no dependence on $\theta$ and $\phi$ 
is considered as shown by Eq. (\ref{eq13}), which depends only on 
$r$.  As a result, all initial states are s-orbitals ($1s$ and $2s$ for 
the Lyman and Balmer series, respectively) with $l = 0$.  However, 
the external EM radiation that is specified by $\omega_o$ and 
$\mathbf {k_o}$ carries the momentum $l = 1$, which makes the 
angular momentum of the electron to be non-zero at its final state.  
Thus, in this model, the final state of Ly-$\alpha$, Ly-$\beta$ and 
Ly-$\gamma$ are $2p$, $3d$ and $4f$ states respectively; none 
of these states has nodes in their radial probability densities as 
shown in Fig. 2.  

Specifically, for Ly-$\alpha$ – one photon from $n = 1$ to $n =2$ 
gives $l = 1$; however, for Ly-$\beta$ – two photons from $n = 1$ 
to $n = 2$ and from $n = 2$ to $n = 3$ give $l = 2$; and similar 
for Ly-$\gamma$.  The reverse process (emission) is also described 
by the NAE, and it is assumed that in this process the electron returns 
to its original state immediately after the jump, or after some period
of time that can be established experimetally (see Sect. 5 for discussion).  
Similar analysis can be done for the Balmer series absorption, and Fig. 4 
shows that none of the resulting states has nodes in their radial probability 
densities. 

Since quantum jumps are non-unitary processes, they cannot be 
described by the SE, which accounts only for unitary processes, thus, the 
NAE is needed.  However, the SE can be used to compute the probability 
of finding possible states of a quantum system before any interaction with 
its surroundings takes place.  Therefore, the fact that the initial and finals 
states for the considered Lyman and Balmer series are in agreement with 
those already known from the solutions to the SE is expected as these are 
the only states available for the electron in the hydrogen atom.  With the 
SE giving the initial and final states, and $\omega_o$ and $\mathbf {k_o}$
corresponding to these states, the NAE describes the resulting quantum 
jumps, and gives the duration time for the jumps as well as it predicts the 
probability distribution in the final state, which depends on the initial state.
Note that the final states would be different for different initial states, and 
that the NAE correctly predicts them.  As this discussion shows, both the 
SE and NAE are needed to describe quantum jumps.

The modern version of Copenhagen interpretation requires that quantum 
jumps are random and abrupt, and that they obey Bohr's rule (e.g., [3-5]). 
The results obtained in this paper based on the NAE are consistent with 
the Copenhagen interpretation as the NAE directly displays the Bohr rule, 
and the considered quantum jumps are random.  However, the difference 
is that while in the Copenhagen interpretation quantum jumps are 
instantaneous, the NAE predicts their duration times to be very short 
but finite, and allows calculating such times, which can be verifiable by 
experiments.       

According to the results presented in Tables 1 and 2, the duration 
time, $T_n$, for the electron's transitions corresponding to the Lyman 
and Balmer series absorptions is finite and of the order of $10^{-16}$ 
$s$ and $10^{-15}$ $s$, respectively.  Thus, the computed duration 
times for the considered quantum jumps are finite, not instantenous 
as originally postulated (e.g., [1-5]); their comparison to experimental 
data is presented and discussed in Sect. 5.  The velocities of the quantum 
transitions given in Tables 1 and 2 are higher for the higher frequencies 
$\omega_o$ (or energies of EM radiation).  These are expected results 
and they imply that the highest velocity would be reached for the electron 
that leaves the hydrogen atom during its ionization - see again Sect. 5 
for more details. 

As briefly described in the Introduction, quantum trajectory theory (QTT) 
based on the stochastic Schrödinger equation (e.g., [20,21,23]) allows 
finding non-real and subjectively real quantum trajectories of individual 
particles that obey the probabilities computed from the SE; this makes 
QTT compatible with the SE.  There are some similarities between QTT 
and the NAE as both describe individual elementary particles.  Thus, 
a modified version of QTT based on the NAE can be developed within
the framework of subjective real quantum trajectories that require 
a detector to observe a quantum system, and the NAE's parameters 
$\omega_o$ and $\mathbf {k_o}$ may play the role of such a 
'detector'; however, such a task is out of the scope of this paper.  
Then, QTT and its modifed version would be complementary to 
each other, similarly to the SE and NAE.

An interesting result is that the periods corresponding to $\omega_o$, 
and the predicted values of the duration times $T_n$ for the Lyman 
and Balmer series given in Tables 1 and 2, respectively, are of similar 
orders.  Nevertheless, the obtained theoretical results are consistent with 
the time-energy uncertainty relation $\Delta t\ \Delta E \geq \hbar /2$,
which still remains a controversial issue in QM (e.g., [38], and references 
therein).  Taking $\Delta t \sim T_n$ and $\Delta E = \hbar \omega_o$,
the relation becomes $T_n\ \omega_o \geq 1/2$, and as the results 
of Tables 1 and 2 demonstrate, it is satisfied for all the theoretically 
predicted values.  It must be noted that the predicted values of 
the duration times $T_n$ are very short, and their experimental 
verification is now considered.

\section{Experimental verification}

Measurements of quantum jumps are difficult because it requires
a time-measuring device that captures the begining of a quantum 
jump.  As a result, in the previous experiments [7-12], it was only 
confirmed that quantum jumps occured randomly and abruptly.
In the experiment performed by Minev et al. [13], the time-scales 
associated with quantum jumps are of an order of microseconds.  
However, these measured time-scales cannot be directly compared 
to the duration times of quantum jumps predicted in this paper. 
The reason is that the measurements in [13] were performed 
with artificial atoms, which are much larger in size than atoms, 
but they have discrete energy levels; was suggested 
that the measured time-scales could be explained by quantum 
trajectory theory (QTT), but it was not explicitly demonstrated.

As the results of Tables 1 and 2 show, $T_n$ computed in this 
paper for the hydrogen atom are about $10$ orders of magnitude 
shorter.  However, using Eqs (\ref{eq5}) and (\ref{eq11}), the 
frequencies $\omega_o$ and $\Omega_n$ can be calculated for 
a Rydberg atom with large values of $n$.  In extreme cases, 
Rydberg states corresponding to $n > 500$ or even $n > 1000$ 
can also be considered (e.g., [41,42], and references therein).  
Simple estimates based on the presented theory demonstrate 
that in order to obtain $T_n \sim 10^{-6}$ $s$, it would 
require Rydberg states with $n > 1000$.  However, it remains 
to be determined whether the prediction made by using such 
extreme Rydberg states are relevant to the experimental data 
reported in [13].

The measurements of quantum jumps in shelved-electron 
experiments (e.g., [7-9], also [43] and references therein) 
demonstrated that individual atoms, after absorbing EM radiation, 
stay in their excited states for periods ranging from a few tenths 
of a second to a few seconds before jumping again; see also [44]
for most recent results in artificial atoms.  In such experiments, 
an electron is temporarily placed in a long-lived, shelved 
(non-fluorescent) state that turns off the atom's fluorescence 
signal; the precise moment when the electron transitions back to 
a fluorescing state is identified as a quantum jump.  Several other 
indepedent experiments (e.g., [10-13]) confirmed this range.  The 
theoretical results presented in this paper do not describe this kind 
of experiments; however, it is suggested that the NAE could be 
used to predict radial probablility densities in spherical shell 
volume elements for the electron in its shelved (non-radiating) 
state.

Significant progress in absolute timing quantum jumps has been 
made since attosecond spectroscopy was applied to photoionization
(e.g., [45] and references therein).  Chronoscope measurements 
of the times involved in the photoelectric effect resulted in the 
duration of the primary photoexcitation process to be of the 
order of $10^{-17}$ $s$ [46,47].  On the other hand, estimates
based on the Franck-Condon principle give the duration times for 
quantum jumps to be of the order of $10^{-15}$ $s$ [48]. 
A specific prediction based on the assumption that quantum 
jumps occur as a result of a resonance of the atomic electron 
with with the modes of the zero-point radiation field of 
Compton's frequency, gives $5 \cdot 10^{-17}$ $s$ [46].
The duration times for different quantum jumps corresponding 
to the Lyman and Balmer series given in Tables 1 and 2 lie 
well within the range of the empirical and theoretical 
evaluations reported in [45-48].  

Measurements performed on a neutral helium atom, which
was hit by a high energy laser pulse, showed that one of the 
electrons was ripped out of the atom causing helium to be 
ionized.  The reported experimental results demonstrated that 
the process occured on time-scale of attoseconds (e.g., [45-47]).  
For the hydrogen atom considered in this paper, the frequency of 
EM radiation required for the atom's ionization is $\omega_o = 
2.06 \cdot 10^{16}$ $s^{-1}$.  In the developed theory based 
on the NAE, the electron's final state must be specified.  Taking 
$n = n_{f} \sim 1000$, Eq. (\ref{eq11}) gives the duration time 
to be of an order of attosecond.  However, for the fully ionized 
atom $n = n_{f}\rightarrow \infty$, which is not allowed by the 
theory that is only valid if the final states of the electron are finite.   

The theoretical results presented in this paper give the radial 
probablility densities in spherical shell volume elements for the 
Lyman and Balmer series absorptions.  According to the experimental 
results reported by the {\it NIST Physical Measurement Laboratory}
[49], the atomic lifetimes of Ly-$\alpha$, Ly-$\beta$ and Ly-$\gamma$
are $4.7 \cdot 10^{-8}$ $s$,  $0.56 \cdot 10^{-8}$ $s$ and  $0.13 \cdot 
10^{-8}$ $s$, respectively.  However, for $H\alpha$, $H\beta$ and 
$H\gamma$, the same source gives $0.44 \cdot 10^{-8}$ $s$, $0.084 
\cdot 10^{-8}$ $s$ and $0.025 \cdot 10^{-8}$ $s$, respectively.
Comparison of these atomic lifetimes to the results of Tables 1 and 2,
it is seen that that the duration times $T_n$ are eight or nine orders 
of magnitude shorter than the the corresponding lifetimes.  Therefore, 
it is suggested that the computed probabilty densities may be directly 
observed by using a quantum microscope similar to that designed by 
researchers [50] who used it to measure the orbital structure of Stark 
states in an excited hydrogen atom.  There are other methods to measure 
quantum phenomena [51-53], but some of them may not be suitable to 
observe single orbitals in a hydrogen atom [53].

There are three main experimental limitations that might affect the 
detection of quantum jumps as predicted by the NAE.  First, the uncertainty 
principle would affect the parameters $\omega_o$ and $\mathbf {k_o}$, 
and this uncertainty would affect the calculations of time scales and probability 
densities, introducing a degree of fuzziness.  Second, measurement backaction
caused by EM radiation used in a quantum microscope would alter the electron 
state, which is not fully captured by the NAE.  Third, apparatus limitations 
(e.g., resolving power) lead to uncertainties in the determination of the 
electron's position, which would affect the accuracy of the probability 
densities obtained from the NAE. 

Having demonstrated that the NAE gives a solution to the quantum 
measurement problem [36], and that it alllows finding the duration 
times for quantum jumps as well as their resulting probability 
densities, as this paper shows, the NAE may also be used to develop 
new quantum-based technologies or improve/modify the currently 
known technologies used in cryptographic systems and in secure 
direct communications (e.g., [54]).

\section{Conclusions}

In this paper, a new asymmetric equation, which is 
complementary to the Schr\"odinger equation, is used 
to develop a theory of quantum jumps.  The main 
advantage of this new theory is that it explicitly displays
Bohr's rule for quantum jumps and it allows for the 
theory to be non-unitary.  The solutions to the new 
equation are used to determine the time-scales of 
quantum jumps, and to calculate the radial probability 
density of finding the electron after its quantum jump.  
The obtained solutions are applied to the Lyman and 
Balmer series.

The theoretical results obtained in this paper are qualitatively 
in agreement with the experimental results given by Minev et 
al. [13], who showed that quantum jumps come at random 
times, but once they come, the evolution of each completed 
jump is a continuous and coherent physical process that 
takes place in a finite time.  The specific values of the duration 
times of the considered quantum jumps in the Lyman and 
Balmer series are predicted to be in the range of $10^{-16} 
- 10^{-15}$ $s$.  These values cannot be directly compared 
to the time-scales measured in [13] because they were 
performed on macroscopic artificial atoms.  However, the 
time-scales predicted in this paper lie well within the range 
of the empirical and theoretical evaluations reported in 
[43-46].   

In accordance with the experimental results [6-13], the 
theoretically predicted radial probability densities are 
considered to last for periods ranging from a few tenths 
of a second to a few seconds; the experiments showed 
that after such times, the electron returns to its original 
orbital.  Thus, it is suggested that the presented radial 
probability densities be observed experimentally by 
using a quantum microscope similar to that designed 
in [50], or other methods [50-53] suitable to observe 
single orbitals in a hydrogen atom.\\

\noindent 
{\bf Acknowledgment:} The author thanks two anonymous 
reviewers for valuable comments and suggestions that allow 
me to improve the original version of this paper.  I am also 
indebted to Reviewer 2 for correcting some historical facts 
about quantum jumps presented in Introduction.  Special 
thanks to Dora Musielak for reading the earlier version of
this manuscript and suggesting improvements in its 
presentation of the results.\\

\noindent 
{\bf Funding:} No funding was received for this work.\\

\noindent 
{\bf Conflict of Interest/Competing Interest:} The author 
declares no conflict of interest.\\

\noindent 
{\bf Data availability statement:} All the data generated 
in this research is available directly in the paper.\\


\begin{thebibliography}{}
\bibitem{1} Bohr, N. (1913) On the constitution of atoms and molecules. 
                     Phil. Mag.,  26, 1       
\bibitem{2} Einstein, A. (1916) On the quantum theory of radiation, CPAE, 
                   The collected papers of Albert Einstein, Edited by J. Stachel et al., 
                   Vols. 1-12. Princeton: Princeton University Press, 1987–2010; Vol. 6, Doc. 38.
\bibitem{3} Schr\"odinger, E.  (1952) Are there quantum jumps? 
                     British J. Phil. Sci.  3, 109    
\bibitem{4} Baggott, J.  (1992) The Meaning of Quantum Theory; Oxford Uni. Press,
                   Oxford, U.K.
\bibitem{5} Merzbacher, E.  (1998) Quantum Mechanics; Wiley \& Sons, Inc.: New York, NY, 
                  USA
\bibitem{6} House, J.E.  (2017) Fundamentals of Quantum Mechanics; Academic Presss: 
                  Cambridge, MA, USA, 2017.
\bibitem{7} Nagourney, W.     Sandberg, J.     Dehmelt, H.  (1986) Shelved optical electron amplifier: 
                  Observation of quantum jumps.    Phys. Rev. Lett.  56, 2797   
\bibitem{8} Sauter, T.     Neuhauser, W.     Blatt, R.     Toschek, P.E.  (1986) Observation of quantum 
                   jumps.  Phys. Rev. Lett.  57, 1696
\bibitem{9} Bergquist, J.C., Hulet, R.J., Itano, W.M. and Wineland, D.J.  (1986) Observation 
                  of quantum jumps in a single atom.  Phys. Rev. Lett.  57, 1699 
\bibitem{10} Basché, Th.,  Kummer, S., Br\"auchle, C.  (1995)  Direct spectroscopic observation of 
                   quantum jumps of a single molecule.  Nature  373, 132 
\bibitem{11} Gleyzes, S.,  Kuhr, S.,  Guerlin, C.,  Bernu, J., Del\'eglise, S.,  Hoff, U.B.,  Brune, M.,     
                    Raimond, J.-M. and Haroche, S.  (2007) Quantum jumps of light recording the birth 
                    and death of a photon in a cavity.  Nature  446, 297
\bibitem{12} Guerlin, C.,  Bernu, J.,  Del\'eglise, S., Sayrin, C.,  Gleyzes, S.,  Kuhr, S.,  Brune, M.     
                    Raimond, J.-M. and Haroche, S. (2007)  Progressive field-state collapse and quantum 
                    non-demolition photon counting.  Nature  448, 889
\bibitem{13} Minev, Z.K.,  Mundhada, S.O.,  Shankar, S.,  Reinhold, P.,  Guti\'errez-Jáuregui, R.     
                    Schoelkopf, R.J., Mirrahimi, M.,  Carmichael, H.J. and Devoret, M.H.  (2019) To catch 
                    and reverse a quantum jump mid-flight.  Nature  570, 200
\bibitem{14} Brewer R.G. and Schenzle, A. (1987) in Laser Spectroscopy VIII, Proc. Eight
                   Int. Conf., {\AA}re. Sweden. June 22-26, 1987, Eds. W. Persson and
                   S. Svanberg, Springer-Verlag, Heidelberg-Berlin, 108                
\bibitem{15} Nienhuis, G.  (1987) in Laser Spectroscopy VIII, Proc. Eight Int. Conf., 
                   {\AA}re, Sweden. June 22-26, 1987, Eds. W. Persson and S. Svanberg, 
                   Springer-Verlag, Heidelberg-Berlin, 112
\bibitem{16} Zoller, P.,  Marte, M. and Walls, D.P.  (1987) Quantum jumps in atomic systems.
                      Phys. Rev. A  35, 198
\bibitem{17} Cook, R.J.  (1988) What are quantum jumps? 
                       Phys. Scripta  T21, 49    
\bibitem{18} Cohen-Tannoudji, C., Zambon B. and Arimondo, E.  (1993) Quantum-jump 
                    appraoch to dissipative processes.  J. Opt. Soc. Am. B 10, 2107    
\bibitem{19} Mabuchi, H.  and  Zoller, P.  (1996) Inversion of quantum jumps in quantum 
                    optical systems under continuous observation.  Phys. Rev. Lett.  76, 3108  
\bibitem{20} Dalibard, J.  and  Castin, Y.  (1992) M$\oslash$lmer, K.  Wave-function approach 
                    to dissipative processes in quantum optics.  Phys. Rev. Lett.  68, 580   
\bibitem{21} Carmichael, H.J.  (1993) An Open System Approach to Quantum Optics, 
                    Springer-Verlag, Berlin  
\bibitem{22} Gisin, N.  and  Percival, I.  (1993) Quantum state diffusion, localization and 
                   quantum dispersion entropy.   J. Phys. A  26, 2233  
\bibitem{23} Wiseman, H.M.  (1996) Quantum trajectories and quantum measurement theory.
                    Quantum Semiclass. Opt.  8, 205    
\bibitem{24} Brun, T.A.  (2002) A simple model of quantum trajectories.  Am. J. Phys.  70, 719   
\bibitem{25} Fr\"ohlich, J.,  Gang, Z.  and  Pizzo, A.  (2024) A theory of quantum jumps.
                      arXiv:2404.10460v3 [quant-ph]  2024,  12 May
\bibitem{26} Del\'eglise, S.,  Dotsenko, I.,  Sayrin, C., Bernu, J.,  Brune, M.,  Raimond, J.-M.     
                    and Haroche, S.  (2008) Reconstruction of non-classical cavity field states with snapshots 
                    of their decoherence.  Nature  455, 510    
\bibitem{27} Sayrin, C.,  Dotsenko, I.,  Zhou, X.,  Peaudecerf, B.,  Rybarczyk, T.,  Gleyzes, S.;
                   Rouchon, P.,   Mirrahimi, M.,  Amini, H,  Bernu, J.,  Brune, M.,  Raimond, J.-M.     
                   and Haroche, S.  (2011) Real-time quantum feedback prepares and stabilizes photon 
                   number states.  Nature  477, 73       
\bibitem{28} Sun, L.,  Petrenko, A.,  Leghtas, Z.,  Vlastakis, B.,  Kirchmair, G.,  Sliwa, K.M.,  Narla, A.,
                    Hatridge, M.,  Shankar, S., Blumoff, J.,  et al.  (2014) Tracking photon jumps with repeated 
                    quantum non-demolition parity measurements.  Nature  511, 444   
\bibitem{29} Ofek, N., Petrenko, A., Heeres, R., Reinhold, P.,  Leghtas, Z., Vlastakis, B., Liu, Y.,     
                    Frunzio, L.,  Girvin, S.M.  and  Jiang, L.  (2016) Extending the lifetime of a quantum 
                    bit with error correction in superconducting circuits.  Nature  536, 441    
\bibitem{30} Musielak, Z.E.  (2021)  New equation of nonrelativistic physics and theory of 
                    dark matter.    Int. J. Mod. Phys. A  36, 2150042.
\bibitem{31} Bargmann, V.  (1954) On unitary ray representations of continuous groups. 
                    Ann. Math.  59, 1
\bibitem{32} Levy-Leblond, J.-M. (1967) Nonrelativistic particles and wave equations. 
                      Comm. Math. Phys.  6, 286
\bibitem{33} Levy-Leblond, J.-M. (1969) Group-theoretical foundations of classical mechanics:
                   The Lagrangian gauge problem.  J. Math. Phys. 12, 64
\bibitem{34} Musielak, Z.E. and Fry, J.L. (2009) Physical theories in Galilean space-time and the 
                   origin of Schrödinger-like equations. Ann. Phys.  324, 296
\bibitem{35} Musielak, Z.E. and Fry, J.L. General dynamical equations for free particles and their 
                    Galilean invariance.  Int. J. Theor. Phys.  48, 1194
\bibitem{36} Musielak, Z.E.  (2024) A solution to the quantum measurement problem.    Quantum 
                    Reports  6, 522 
\bibitem{37} Wigner, E.P. (1939) On unitary representations of the inhomogeneous Lorentz group. 
                     Ann. Math.  40, 149 
\bibitem{38} In\"onu, E.  and  Wigner, E.P.  (1952) Representations of the Galilei group.    
                    Nuovo C.  9, 705 
\bibitem{39} Kim, Y.S.  and  Noz, M.E.  (1986) Theory and Applications of the Poincar\'e Group; 
                    Reidel: Dordrecht, The Netherlands
\bibitem{40} Busch, P.  (2008) The time-energy uncertainty relationship, in Time in Quantum 
                    Mechanics, Eds. Muga, J.G.     Sala Mayato, R.     Equsquiza, \'I.L., Springer-Verlag,
                    Berlin, Heidelberg, Germany       
\bibitem{41} Saffman, M.,  Walker, T.G. and Molmer, K.  (2010) Quantum information with Rydberg atoms. 
                       Rev. Mod. Phys.  82, 2313    
\bibitem{42} Noordam, L.D.  and  Jones, R.R. (1997) Probing Rydberg electron dynamics.    
                   J. Mod. Opt.  44, 2515
\bibitem{43} Itano, W.M.,  Bergquist, J.C. and  Wineland, D.J.  (2015) Early observations of macroscopic quantum 
                    jumps in single atoms.  Int. J. Mass Spectrom.  377, 403
\bibitem{44} Cottet, N.,  Xiong, H.,  Nguyen, L.B., Lin, Y.-H. and Manucharyan, V.E.  (2021) Electron shelving of a 
                    superconducting artificial atom.  Nature Comm.  Article number: 6383
\bibitem{45} Pazourek, A.,  Nagele, S. and Burgd\"orfer, J.  (2015) Attosecond chronoscopy of photoemission.
                       Rev. Mod. Phys.  87, 765   
\bibitem{46} Ossiander, M., Siegrist, F.,  Shirvanyan, V., R. Pazourek, R.,  Sommer, A.,     
                    Latka, T.,  Guggenmos, A.,  Nagele, S.,  Feist, J.,  Burgd\"orfer, J.  et al.  (2017) Attosecond
                    correlation dynamics.    Nature Phys.  13, 280    
\bibitem{47} Ossiander, M.,  Riemensberger, J.,  Neppl, S.,  Mittermair, M.,  Sc\"affer, M., Duensing, A.
                    Wagner, M.S.,  Heider, R.,  Wurzer, M.,  Gerl, M.  et al.  (2018) Absolute timing of the photoelectric 
                    effect.  Nature  561, 374 
\bibitem{48} de la Pena, L.,  Cetto, A.M.  and Vald\'es-Hern\'andez, A.  (2020) How fast are qunatum jumps?
                       Phys. Lett. A  384, 126880
\bibitem{49} Ralchenko, Y.  (2022) Atomic Spectroscopy - Atomic Lifetimes.    NIST Physical Measuerment 
                    Laboratory; https://www.nist.gov/pml/atomic-spectroscopy-compendium-basic
                    -ideas-notation-data-and-formulas/atomic-spectroscopy-atomic
\bibitem{50} Stodolna, A.S.,  Rouz\'ee, A.,  L\'epine, F., Cohen, S., Robicheaux, F., Gijsbertsen, A.,     
                    Jungmann, J.H.,  Bordas, C.  and  Vrakking, M.J.J.  (2013) Hydrogen Atoms under Magnification: 
                    Direct Observation of the Nodal Structure of Stark States.  Phys. Rev. Let.  110, 213001
\bibitem{51} McCarthy, I.E. and  Weigold, E.  (1988) Wavefunction mapping in collision experiments. 
                       Rep. Progr. Phys.  51, 299
\bibitem{52} Itatani, J.,  Levesque, J.,   Zeidler, D.,  Niikura, H.,  P\'epin, H.,  Kieffer, J.C.,  Corkum, P.B. and     
                    Villeneuve, D.M.  (2004) Tomographic imaging of molecular orbitals.  Nature  432, 867
\bibitem{53} Shafir, D.,  Mairesse, Y.,  Villeneuve, D.M.,  Corkum, P.B.  and  Dudovich, N.  (2009) Atomic 
                    wavefunctions probed through strong-field light–matter interaction.  Nat. Phys.  5, 412
\bibitem{54} Pan, D.,  Long, G.L.,  Yin, L.,  Sheng, Y.B.,  Ruan, D.,  Ng S.X.,  Lu, J.  and Hanzo, L. 
                    (2024) The evolution of quantum secure direct communication: on the road to the qinternet.    
                    IEEE Comm. Surv. Tutor.  26, 1898


\end{thebibliography}
\end{document}